\documentclass[aps,prd,eqsecnum,
epsf,nofootinbib]{revtex4}
\usepackage{amsfonts,latexsym,eucal,amsmath,amssymb,times,mathrsfs}
\usepackage[dvips]{graphicx}
\usepackage[usenames]{color}

\newcommand{\bm}[1]{\hbox{\boldmath{$#1$}}}
\newcommand{\sbm}[1]{\hbox{\boldmath{\scriptsize$#1$}}}

\newcommand{\Mp}{M_{\rm pl}}
\newcommand{\dd}{{\rm d}}

\newcommand{\sR}{{^s\!R}}

\newcommand{\gR}{{^g\!R}}
\newcommand{\gz}{{^g\!\zeta}}
\newcommand{\Approx}{\stackrel{\rm  IR}{\approx}}
\newcommand{\cL}{{\cal L}}

\newcommand{\gauge}{{\em gauge} }
\newcommand{\cv}{\{\zeta,\pi\}}
\newcommand{\cvt}{\{\tilde{\zeta},\tilde{\pi}\}}

\begin{document}

\thispagestyle{empty}


\title{Strong restriction on inflationary vacua from the local \gauge
invariance I :\\
Local \gauge invariance and infrared regularity }
\date{\today}
\author{Takahiro Tanaka$^{1}$}
\email{tanaka_at_yukawa.kyoto-u.ac.jp}
\author{Yuko Urakawa$^{2}$}
\email{yurakawa_at_ffn.ub.es}
\affiliation{\,\\ \,\\
$^{1}$ Yukawa Institute for Theoretical Physics, Kyoto university,
  Kyoto, 606-8502, Japan\\
$^{2}$ Departament de F{\'\i}sica Fonamental i Institut de Ci{\`e}ncies del Cosmos, 
Universitat de Barcelona,
Mart{\'\i}\ i Franqu{\`e}s 1, 08028 Barcelona, Spain}



\begin{abstract}
The primordial perturbation is widely accepted to be generated through
 the vacuum fluctuation of the scalar field which drives inflation. It is,
 however, not completely clear what is the natural vacuum in the inflationary
 universe particularly in the presence of non-linear interactions. 
In this series of papers, we will address this issue, focusing on the condition 
required for the removal of the divergence from the infrared (IR)
 contribution to loop diagrams. We show that requesting the \gauge
 invariance in the local observable universe guarantees the IR regularity
 of the loop corrections beginning with a simple initial state. In our previous works, the
 IR regularity condition was discussed using the slow roll
 expansion, which restricts the background evolution of the inflationary
 universe. We will show more generally that requesting the \gauge invariance/the IR
 regularity leads to non-trivial constraints on the allowed quantum states.  
\end{abstract}


\maketitle


\section{Introduction}   \label{Sec:Intro}
It is widely accepted that primordial curvature perturbations 
originate from the vacuum fluctuation of the inflaton field 
in the inflationary universe. 
Different choices of quantum states lead to different
statistics of the primordial fluctuation. Among various options of
vacuum states, we usually choose the adiabatic vacuum. 
A free scalar field can be understood as a set of independent 
harmonic oscillators. Taking the adiabatic vacuum looks reasonable,
because each oscillator tends to have 
an approximately fixed frequency as the time scale becomes shorter
than that of the cosmic expansion. However, once we start to 
take into account the self-interaction of the field, the reasoning 
of choosing the adiabatic vacuum becomes obscure. 
Therefore it is an interesting question whether some physical
requirement prohibits taking an arbitrary quantum state or not. This is
the question we address in this series of papers. 

The key idea is to focus on fluctuations that we can measure in observations. Since the region we
can observe is limited to a portion of the whole universe, we need to take into 
account the local property of the observable fluctuations. In case we
can get access to information only on a portion of the 
whole universe, there appear additional degrees of freedom to choose
boundary conditions in fixing the coordinates. As shown in our recent
publications~\cite{IRsingle, IRmulti, IRgauge_L, IRgauge, IRNG,
IRgauge_multi} and will be briefly described later, these degrees of freedom in the boundary conditions can
be thought of as residual degrees of freedom in fixing the
coordinates of the local observable universe. The observable
fluctuations should be insensitive to the 
coordinate choice in the local observable
universe. We showed that removing this ambiguity 
significantly affects the infrared (IR) behaviour of
primordial fluctuations. In the conventional gauge-invariant
perturbation, in which this ambiguity is not taken care of, it
is widely known that loop corrections of massless fields such as the inflaton
diverge because of the IR contributions~\cite{Boyanovsky:2004gq, Boyanovsky:2004ph, Boyanovsky:2005sh,
Boyanovsky:2005px, Tsamis:1996qm, Tsamis:1996qq, Onemli:2002hr, Brunier:2004sb, Prokopec:2007ak,
Sloth:2006az, Sloth:2006nu, Seery:2007we, Seery:2007wf, Urakawa:2008rb,
Adshead:2008gk, Cogollo:2008bi, Rodriguez:2008hy, Seery:2009hs,
Gao:2009fx, Bartolo:2010bu, Seery:2010kh,
Kahya:2010xh}. One may expect that performing the quantization 
after we remove these residual
coordinate degrees of freedom in the local universe can cure the
singular behaviour of the IR contributions. 
However, these residual coordinate degrees of freedom 
are not present if we deal with the whole universe 
including the spatial infinity, since what we call the residual coordinate 
transformation diverges at spatial infinity. 
In this sense, if we think of the whole universe, they
are not gauge degrees of freedom
In Ref.~\cite{IRsingle}, we imposed the boundary condition at a finite distance 
to remove the degrees of freedom in choosing coordinates, 
but the quantization (or equivalently setting the initial 
quantum sate) was performed considering the whole spatial 
section of the universe. 
In that analysis we concluded that there is no IR divergences 
irrespective of the initial quantum state. 
However, there were several overlooked points. 
After we set the initial quantum state at a 
finite initial time, we need to translate it into the language 
written solely in terms of the local quantities. 
Namely, the Heisenberg operators corresponding to the 
perturbation variables should be transformed into the ones that 
satisfy given boundary conditions using an appropriate 
residual spatial coordinate transformation. 
Since the non-linear part of this transformation is not guaranteed 
to be IR regular, it can produce IR divergences.

If we send the initial time to the past infinity, one may think 
that this problem may disappear because 
our observable region in comoving coordinates becomes 
infinitely large in this limit. 
However, this argument is not so obvious. If 
we think of, for instance, the conformal diagram of 
de Sitter space, the causal past of an observer never  
covers the whole region of a time constant slice in flat chart. 
Furthermore, in this case we also need to care 
about the secular 
growth of fluctuation. In Ref.~\cite{IRsingle}, we discussed the absence of the secular
growth, focusing only on modes beyond the horizon length
scale. To complete the discussion, we also need to include the
vertexes which contain modes which are below the Hubble scale, because these vertexes
may yield the secular growth through correlations between sub horizon
modes and the super horizon modes.

Later in Refs.~\cite{IRgauge_L, IRgauge}, 
leaving the residual coordinate degrees of
freedom unfixed, we calculated quantities that are 
invariant under the residual coordinate transformation 
in the local universe under the so called slow-roll approximation. 
Then, we derived the conditions on
the initial states that guarantee the IR regularity. 
These conditions are the ones that guarantee the absence 
of additional IR divergences originating from the coordinate 
transformation mentioned above. 
However, the physical meaning of these conditions  
is not transparent from the derivation in Refs.~\cite{IRgauge_L, IRgauge}.

In this paper, as in the conventional perturbation theory
and also as in Refs.~\cite{IRgauge_L, IRgauge}, 
we will perform the quantization over the whole universe. 
Then, calculating observable quantities, which are 
invariant under the residual coordinate transformation 
in the local universe,  
we will derive the necessary conditions 
for the initial quantum state to be free from IR divergences 
more generally, removing the limitation to 
the slow-roll approximation. 
We will also show that the IR 
regularity conditions can be thought of as the
conditions that request the equivalence between two systems 
related by means of the dilatation transformation of spatial
coordinates. In this paper, we will consider a case in which 
the interaction is turned on at a finite initial time. In the succeeding
paper, we will discuss the case when the initial time is sent to the
past infinity. This paper is organized as follows. In
Sec.~\ref{Sec:P}, we first show a symmetry of the system, which plays an
important role in our discussion. Then, following Ref.~\cite{IRgauge_L,
IRgauge}, we construct an operator which remains invariant under 
the residual coordinate transformation. In Sec.~\ref{Sec:Fin}, we
address the conditions on the initial states. In Sec.~\ref{Sec:Con}, we
will summarize the results of this paper and will outline future issues.

\section{Preparation}   \label{Sec:P}
In this section, we will introduce basic ingredients to calculate
observable fluctuations. In this paper, we consider a standard single
field inflation model whose action takes the form  
\begin{eqnarray}
 S = \frac{\Mp^2}{2} \int \sqrt{-g}~ [R - g^{\mu\nu}\phi_{,\mu} \phi_{,\nu} 
   - 2 V(\phi) ] \dd^4x~, \label{Exp:action} 
\end{eqnarray}
where $\Mp$ is the Planck mass and $\phi$ is a dimensional scalar field
divided by $\Mp$. In Sec.~\ref{SSec:Sym}, we show a
symmetry of this system, which plays a key role in our argument. In
Sec.~\ref{SSec:Quantization}, after we introduce a variable which
preserves invariance regarding the coordinate choice in the local
universe, we provide a way of quantizing the system. 

\subsection{Symmetry of the system}  \label{SSec:Sym}
To fix the time slicing, we
adopt the uniform field gauge $\delta\phi=0$. Under the ADM metric
decomposition, which is given by
\begin{eqnarray}
 \dd s^2 = - N^2 \dd t^2  + h_{ij} (\dd x^i + N^i \dd t) (\dd x^j + N^j
  \dd t)~, \label{Exp:ADMmetric}
\end{eqnarray}
we take the spatial metric $h_{ij}$ as
\begin{align}
 & h_{ij} = e^{2(\rho+\zeta)} \left[ e^{\delta \gamma} \right]_{ij}\,,
\end{align}
where $e^\rho$ denotes the background scale factor, $\zeta$ is the
so-called curvature
perturbation and $\delta \gamma_{ij}$ is traceless:
\begin{align}
 \delta {\gamma^i}_i=0\,,
\end{align}  
where the spatial index was raised by $\delta^{ij}$.
As spatial gauge conditions we impose 
the transverse conditions on $\delta \gamma_{ij}$:
\begin{align}
 & \partial_i \delta \gamma^i\!_j=0\, . 
\label{TTgauge}
\end{align} 
In this paper, we neglect the vector and tensor perturbations. The
tensor perturbation, which is a massless field, can also contribute to
the IR divergence of loop corrections. We will address this issue in our
future publication.

Solving the Hamiltonian and momentum constraints, we can derive the
action which is expressed only in terms of the curvature
perturbation~\cite{Maldacena2002}. Since the spatial metric
is given in the form $ e^{2\rho}e^{2\zeta} \dd \bm{x}^2$, we naively
expect that the explicit
dependence on the spatial coordinates
appear in the action for $\zeta$ only in the form of the physical
distance $e^\rho e^\zeta \dd \bm{x}$. We first examine this property.

Using the Lagrangian density in the physical
coordinates ${\cal L}_{phys}$, we express the action as
\begin{align}
 S &  = \int \dd t\, \dd {\bm x}\, {\cal L}(x) = \int \dd t\, \dd {\bm x}\, e^{3(\rho+\zeta)} {\cal L}_{phys} (x)\,.
\end{align}
We can confirm that the Lagrangian density ${\cal L}_{phys}$ is composed of
terms which are in a covariant form regarding a spatial coordinate transformation such as
$h^{ij} \partial_i N_j$ and $h^{ij} N_i \partial_j \zeta$.
Using 
\begin{align}
 & \dd \check{x}^i = e^{\rho+\zeta} \dd x^i \,, \qquad
 \partial/\partial \check{x}^i = e^{-(\rho+\zeta)} \partial/\partial x^i \,, \qquad  \check{N}_i  =
 e^{-(\rho+\zeta)} N_i \,,
\end{align}
we can absorb the curvature perturbation $\zeta$ without differentiation
which appears from the spatial metric, for instance, as 
\begin{eqnarray}
 h^{ij} \partial_i N_j =   \delta^{ij} e^{-2(\rho+\zeta)} \partial_i N_j =
  \delta^{ij}  \left( \frac{\partial}{\partial \check{x}^i}
			  \check{N}_j + \check{N}_j \frac{\partial}{\partial
			  \check{x}^i} \zeta \right)\,.
\end{eqnarray}
After a straightforward repetition, the Lagrangian density can be recast into
\begin{align}
 S & = \int \dd t\, \dd {\bm x}\, e^{3(\rho+\zeta)} {\cal L}_{phys} [
  {\cal D} \zeta, N,\, \check{N}_i] \,. \label{Exp:L_ADM}
\end{align}
Here, to stress the fact that all $\zeta$s which remain unabsorbed are
associated with differentiation 
$\partial_t$ or $\partial/\partial \check{x}^i$, we introduced
${\cal D}$ which denotes differentiations in general. 
Using this expression
of the action, the Hamiltonian constraint is given by
\begin{align}
 & {\cal C}  [ {\cal D}\zeta,
 N,\, \check{N}_i] := \frac{\delta {\cal
 L}_{phys}}{\delta N} = e^{-3(\rho+\zeta)}  \frac{\delta {\cal L}}{\delta N} =0\,,
\end{align}
and the momentum constraints are given by
\begin{align}
 &  {\cal C}_ i [ {\cal D}\zeta,
 N,\, \check{N}_i] := \frac{\delta {\cal
 L}_{phys}}{\delta \check{N}^i} =  e^{-2(\rho+\zeta)}  \frac{\delta {\cal L}}{\delta N^i} =0\,.
\end{align}
Perturbing these constraints, we can show that the Lagrange multipliers
$N$ and $\check{N}_i$ are given by solutions of the elliptic type equations:
\begin{align}
 &  \left( \frac{\partial}{\partial \check{\bm x}} \right)^2 N =  f  [
  {\cal D}\zeta(x)]\,,
 \qquad 
  \left( \frac{\partial}{\partial \check{\bm x}} \right)^2 \check{N}_i =  f_i [
  {\cal D}\zeta(x)]\,. \label{Eq:Constraint}
\end{align}
To address the system of the whole universe, the integration in the
action is assumed to be taken over the whole universe. Then, by assuming the
regularity at the spatial infinity, these Poisson equations can be
uniquely solved as  
\begin{align}
 & N = N [{\cal D}\zeta(x)]\,, \qquad \qquad \check{N}_i =
 \check{N}_i[{\cal D}\zeta(x)]\,.
\end{align}
Substituting these expressions of $N$ and $\check{N}_i$ into the action
(\ref{Exp:L_ADM}), we can show that the action takes the form:
\begin{eqnarray}
 S = \int \dd t\, \dd^3 \check{\bm{x}}\,   {\cal L}_{phys} [{\cal
  D}_{\dd \check{\sbm{x}}}\zeta(x)] \,.  \label{Exp:S2}
\end{eqnarray}
Here, we put the subscript $\dd \check{\bm{x}}$ on ${\cal D}$ to specify the
spatial coordinates used in ${\cal D}$. The
equation (\ref{Exp:S2}) explicitly shows that the action for $\zeta$
only depends on the physical spatial distance.

Now we are ready to show the dilatation symmetry of the system, which
plays a crucial role in discussing the IR regularity. (See also the
discussions in \cite{Giddings:2010nc,Byrnes:2010yc,Gerstenlauer:2011ti,Giddings:2011zd,
	Giddings:2011ze, Senatore:2012nq, Pimentel:2012tw}.) Changing 
the coordinates in the integral from 
$\bm{x}$ to $e^{-s} \bm{x}$ with a constant parameter
$s$, we can rewrite the action (\ref{Exp:S2}) as 
\begin{align}
 S  &=  \int \dd t\, \dd^3 \bm{x}\, e^{3(\rho+\zeta)}  {\cal L}_{phys} [{\cal
  D}_{e^{\rho+\zeta} \dd \sbm{x}}\zeta(x)] =  \int \dd t\, \dd^3 \bm{x}\, e^{3(\rho+\zeta-s)}  {\cal L}_{phys} [{\cal
  D}_{e^{\rho+\zeta-s} \dd \sbm{x}}\zeta(t,\, e^{-s}\bm{x})] \,.
\end{align}
This implies that the action for $\zeta$ possesses the dilatation
symmetry 
\begin{align}
 & S = \int \dd t\, \dd^3\! {\bm x}\,  {\cal L}\left[\zeta(x) \right] = 
 \int \dd t\, \dd^3\! {\bm x}\, {\cal L}\left[ \zeta(t,\, e^{-s}\bm{x})
 - s \right]\,. \label{Exp:Dsym}
\end{align}
As long as we consider a theory which preserves the three-dimensional
diffeomorphism invariance, the system preserves the dilatation symmetry. 
It is because the above-mentioned dilatation symmetry can be thought of as
one of the spatial coordinate transformations. This
symmetry is discusses also in Refs.~\cite{Creminelli:2012ed, Hinterbichler:2012nm}.

To make use of the dilatation symmetry, we introduce another set of
canonical variables than $\zeta(x)$ and its conjugate momentum $\pi(x)$. 
We can show that
\begin{align}
 & \tilde{\zeta}(x):= \zeta(t,\, e^{-s}\bm{x}) \,, \qquad \quad
 \tilde{\pi}(x):= e^{-3 s}\pi(t,\, e^{-s}\bm{x}) \label{Def:tilde}
\end{align}
satisfy the canonical commutation relations as well as $\zeta(x)$ and
$\pi(x)$. Actually, using the commutation relations for $\zeta(x)$ and $\pi(x)$,
we can verify  
\begin{align}
 & \left[ \tilde{\zeta}(t,\, {\bm x}),\, \tilde{\pi}(t,\, {\bm y})
 \right] = \left[ \zeta(t,\, e^{-s}{\bm x}),\, e^{-3
 s} \pi(t,\, e^{-s}{\bm y})  \right]
 = e^{-3s} i \delta^{(3)}(e^{-s}({\bm x} -
 {\bm y})) = i \delta^{(3)}({\bm x}- {\bm y})\,,
\end{align}
and also
\begin{align}
 &  \left[ \tilde{\zeta}(t,\, {\bm x}),\, \tilde{\zeta}(t,\, {\bm y})
 \right] =  \left[ \tilde{\pi}(t,\, {\bm x}),\, \tilde{\pi}(t,\, {\bm y})
 \right] = 0\,.
\end{align}

Using Eq.~(\ref{Exp:Dsym}), we can show that the Hamiltonian densities
expressed in terms of these two sets of the canonical variables are
related with each other as
\begin{align}
 \int \dd^3\! {\bm x}\, {\cal H}[\zeta(x),\, \pi(x)] &= \int 
 \dd^3 {\bm x} \left\{ \pi(x) \dot\zeta(x) - {\cal L}[\zeta(x)] \right\} \cr
 & = \int 
 \dd^3\! {\bm x}\, \left\{ e^{- 3s} \pi(t,\, e^{-s}\bm{x})
 \dot\zeta(t,\, e^{-s}\bm{x}) - {\cal L}[\zeta(t,\, e^{-s}\bm{x}) - s] \right\} \cr
 & = \int \dd^3\! {\bm x}\, {\cal H}[\tilde{\zeta}(x)- s,\,
 \tilde{\pi}(x)] =: \int \dd^3\! {\bm x}\, \tilde{{\cal H}}[\tilde{\zeta}(x),\,
 \tilde{\pi}(x)]\,,
\label{Exp:Ht}
\end{align}
where on the second equality, we again changed the coordinates in the
integral as $\bm{x} \to e^{-s} \bm{x}$.
Note that the Hamiltonian density for $\tilde{\zeta}$ and $\tilde{\pi}$ is
given by the same functional as the one for $\zeta$ and $\pi$ with
$\tilde{\zeta}$ shifted by $\tilde{\zeta}-s$.

We define the non-interacting part of the Hamiltonian density for $\tilde{\zeta}$
and $\tilde\pi$ by the quadratic part of ${\cal H}$ in perturbation, 
assuming that $s$ is as small as $\tilde\zeta$ 
and $\tilde\pi$, as follows,
\begin{equation}
 \tilde{{\cal H}}_{0}
[\tilde \zeta(x),\, \tilde \pi(x)] = 
{\cal H}_{0}[\tilde \zeta(x)-s,\, \tilde \pi(x)] \,, \label{Eq:H0}
\end{equation}
where ${\cal H}_0$ is the free part of the Hamiltonian density for
$\zeta$ and $\pi$. Using Eq.~(\ref{Exp:Ht}) with Eq.~(\ref{Eq:H0}), we
can show that the interaction Hamiltonian densities which are defined by
${\cal H}_I:={\cal H}-{\cal H}_0$ for each set of the
canonical variables are related as
\begin{equation}
 \tilde{{\cal H}}_I[\tilde\zeta(x),\,\tilde\pi(x)] =
{\cal H}_I[\tilde\zeta(x)-s,\, \tilde \pi(x)]\,.  \label{Exp:tH}
\end{equation}
Note that the Hamiltonian densities ${\cal H}_I$ and 
$\tilde{{\cal H}}_I$ also take the same functional form
except for the constant shift of $\tilde{\zeta}$. 

\subsection{Residual coordinate transformations and quantization}
\label{SSec:Quantization}
In this subsection, we consider a way to calculate observable
fluctuations. First, we begin with the classical theory. To obtain the
action for $\zeta$ by eliminating the Lagrange multipliers $N$ and $N_i$, we need to solve the Hamiltonian
and momentum constraint equations. As is schematically expressed in
Eqs.~(\ref{Eq:Constraint}), these constraint equations are given by the
elliptic-type equations. Here we note that the region where we can
observationally access 
is restricted to the causally connected region whose
spatial volume is bounded at a finite past. As far as we are
concerned only with the observable region, boundary conditions for
Eqs.~(\ref{Eq:Constraint}) cannot be restricted from the regularity of
the spatial infinity, which is far outside of the observable region. The
degrees of freedom for solutions of $N$ and $N_i$ can be understood as degrees
of freedom in choosing coordinates. Since the time slicing is fixed by
the gauge condition $\delta \phi=0$, there are remaining degrees of
freedom only in
choosing the spatial coordinates. In the following, we refer to such
degrees of freedom which cannot be uniquely specified without the knowledge
from the outside of the observable region as the residual \gauge degrees of
freedom in the local universe. We write the term \gauge in 
the italic fonts, because changing the
boundary conditions for $N$ and $N_i$ in the local region modifies the
action for $\zeta$ obtained after solving constraint equations, while it keeps the action
expressed by $N$, $N_i$, and $\zeta$ invariant. In this sense, the change of the boundary
condition is distinct from the usual gauge transformation, which keeps
the action invariant.

The observable fluctuations should be free from such residual gauge
degrees of freedom. Following Refs.~\cite{IRgauge_L, IRgauge}, we 
construct a genuine \gauge invariant operator, which preserves the \gauge
invariance in the local observable universe. For the construction, we
note that the scalar curvature $\sR$, which transforms as a scalar
quantity for spatial coordinate transformations, become genuinely
\gauge invariant, if 
we evaluate it in the geodesic normal coordinates
span on each time slice. 
The geodesic normal coordinates are 
introduced by solving the spatial
three-dimensional geodesic equation:  
\begin{eqnarray}
 \frac{\dd^2 x_{gl}^i}{\dd \lambda^2} +  {^s \Gamma^i}_{jk} \frac{\dd
  x_{gl}^j}{\dd \lambda} \frac{\dd x_{gl}^k}{\dd \lambda} =0~,
\label{GE}
\end{eqnarray}
where ${^s \Gamma^i}_{jk}$ is the Christoffel symbol with respect to 
the three dimensional spatial metric on a constant time hypersurface and
$\lambda$ is the affine parameter. 
We consider the three-dimensional geodesics whose affine parameter ranges
from $\lambda=0$ to $1$ with the initial ``velocity'' given by
\begin{eqnarray}
 \frac{\dd x^i_{gl}(\bm{x},\lambda)}{\dd \lambda} \bigg\vert_{\lambda=0}= e^{-\zeta(\lambda=0)}
 \bm{x}^i\,. \label{IC}
\end{eqnarray}
Here we put the index $gl$ on the global coordinates, using the simple
notation $\bm{x}$ for the geodesic normal coordinates, which will be
mainly used in this paper. A point $x^i$ in the geodesic normal coordinates is identified 
with the end point of the geodesic, $x_{gl}^i(\bm{x},\lambda=1)$ in the 
original coordinates. 
Using the geodesic normal coordinates $x^i$, we perturbatively 
expand $x_{gl}^i$ as 
$x_{gl}^i= x^i + \delta x^i(\bm{x})$. 
Then, we can construct a genuinely \gauge invariant
variable as 
\begin{align}
 {^g\!R}(x) &:= \sR (t,\, x_{gl}^i(\bm{x})) = 
 \sR (t,\, x^i + \delta x^i (\bm{x})))\,, \label{Def:gR}
\end{align}
As long as the deviation from the FRW universe is kept perturbatively
small, we can foliate the universe by the geodesic normal coordinates.

Next, we quantize the system to calculate quantum correlation functions
which become observable. A straightforward and frequently used way to
preserve the gauge invariance is to eliminate gauge degrees of freedom
by fixing the coordinates completely. In the local observable universe, the complete gauge fixing
requires us to fix the boundary conditions in solving the constraint
equations. However, to quantize the locally restricted system, we need
to abandon several properties which are available in the quantization of
the whole universe. One is that the quantization restricted to a local system
cannot be compatible with the global translation symmetry at least in a
manifest way. Another but related aspect is that basis functions for a mode
decomposition become rather complicated (even if it exists) than the
Fourier modes. To preserve the global translation symmetry manifestly,
we take another way of quantization than the quantization in the completely fixed gauge.

In this paper, we perform quantization in the whole universe with the
infinite spatial volume. The global translation symmetry in the spatial
directions is then manifestly guaranteed in the sense that a shift of
the spatial coordinates $\bm{x}$ to $\bm{x}+\bm{a}$ just changes the
overall phase factor in the Fourier mode by $e^{i\sbm{k}\sbm{a}}$. 
Based on this idea, 
we consider and calculate observable quantities. Since
the gauge invariant variable $\gR$ 
does not include the conjugate momentum of $\zeta$, we can consider 
products of $\gR$ at an equal time 
without the problem of operator ordering. The
$n$-product of $\gR$, {\it i.e.}, $\gR(x_1) \cdots \gR(x_n)$, preserves the
\gauge invariance in the local universe. To calculate the $n$-point functions of $\gR$,
we need to specify the quantum state as well. 
One may think that the quantum state should be also invariant 
under the residual \gauge transformations. 
However, we cannot directly discuss this invariance as 
a condition for allowed quantum states 
in this approach, because the residual
\gauge degrees of freedom are absent when we quantize
fields in the whole universe.

Here we note that even though the operator $\gR$ is not affected by the residual
\gauge degrees of freedom, this does not imply that 
the $n$-point functions of $\gR$ are uncorrelated to the fields 
in the causally disconnected region. 
To explain this aspect more clearly, let us consider the
$n$-point function of $\gR$ at $t=t_f$ whose vertexes are located within
the observable region ${\cal O}_f$. For a later use, we refer to 
the spacetime region which is causally connected to spacetime 
points in ${\cal O}_f$ as the
observable region ${\cal O}$.  
After expanding of the operator $\gR$
in terms of $\zeta_I$, the interaction picture field of $\zeta$, even
if interaction vertexes which affect $\gR(t_f,\, \bm{x})$ are confined 
within the observable region ${\cal O}$, the $n$-point functions of
$\gR$ have correlation to the outside of ${\cal O}$ through 
the Wightman function of $\zeta_I$, which 
can be expressed in the Fourier space as
\begin{align}
 & G^+(x_1,\, x_2)= \langle \zeta_I(t,\, \bm{x}_1) \zeta_I(t,\,
 \bm{x}_2) \rangle 
 = \int \frac{\dd^3 \bm{k}}{(2\pi)^3} e^{i\sbm{k} \cdot
 (\sbm{x}_1-\sbm{x}_2)}  |v_k(t)|^2 \,, 
\end{align}
where $v_k$ is the mode function for $\zeta_I$. Since all the vertexes
are confined within ${\cal O}$, the spatial distance 
$|\bm{x}_1- \bm{x}_2|$ is bounded from above. 
However, the IR modes with 
$k \leq |\bm{x}_1- \bm{x}_2|^{-1}$ are not suppressed and let
the Wightman function $G^+$ diverge for scale-invariant or red-tilted
spectrums.

This long-range correlation becomes the origin of the IR
divergence of loop contributions. Since the observable quantities should
take a finite value, we request the IR regularity of observable
fluctuations, which is achieved only when the quantum state is selected 
so that the long-range correlation is properly isolated from the 
observable quantities. 
In the following section, we will show that requesting the
absence of the IR divergence in fact constrains the quantum state of
the inflationary universe.

\section{IR regularity condition and the residual \gauge invariance}  \label{Sec:Fin}
A simple way to address the evolution of a non-linear system is to
solve the Heisenberg equation perturbatively by assuming that the
interaction is turned on at the initial time. 
In this section, taking this setup, we calculate
the correlation function of $\gR$ up to one-loop order and investigate the IR
behaviour. 

\subsection{Specifying the iteration scheme}
For notational convenience, we introduce the horizon flow functions as 
\begin{align}
 & \varepsilon_0:= \frac{\dot{\rho}_i}{\dot{\rho}},\qquad \quad
 \varepsilon_{n}:= \frac{1}{\varepsilon_{n-1}} \frac{\dd}{\dd \rho}
 \varepsilon_{n-1}~,
\end{align}
with $n\geq 1$, but without assuming these functions are small  
we leave the background inflation model unconstrained. The assumption that
the interaction is turned on at the initial time $t_i$ requests 
\begin{align}
 & \zeta(t_i,\, \bm{x}) = \zeta_I(t_i,\, \bm{x})\,,\qquad \quad
  \pi(t_i,\, \bm{x}) = \pi_I(t_i,\, \bm{x}) \label{IC}
\end{align}
where $\pi_I$ is the conjugate momentum defined from the non-interacting action:
\begin{align}
 &  S_0 = \Mp^2 \int  \dd t \dd^3 \bm{x} \,
e^{3\rho} \dot{\rho}^2 \varepsilon_1
 \biggl[ (\partial_\rho \zeta)^2  -
 \frac{e^{-2\rho}}{\dot{\rho}^2} \! ( \partial \zeta)^2 \biggr] \,,  \label{Exp:S0}
\end{align} 
as $\pi_I:= 2 \Mp^2 e^{3\rho} \varepsilon_1 \dot{\zeta}_I$. Employing
the initial conditions (\ref{IC}), we can relate the Heisenberg picture 
field $\zeta$ to the interaction picture field $\zeta_I$ as  
\begin{align}
 & \zeta (t, \bm{x}) =  U_I^\dagger(t,\,t_i)  \zeta_I(t, \bm{x}) 
U_I(t,\, t_i)\,, \label{Exp:zeta}
\end{align}
where $U_I$ is the unitary operator given by
\begin{align}
 & U_I(t_1,\, t_2) = T \exp \left[- i\int^{t_1}_{t_2} \dd t\, H_I(t)
 \right]\,.
\end{align}
Since the Heisenberg fields are related to the interaction picture
fields by the unitary operator, the canonical commutation
relations for the canonical variables $\zeta$ and  $\pi$ guarantee
those for $\zeta_I$ and $\pi_I$ as 
\begin{align}
 &  \left[ \zeta_I(t,\, \bm{x}),\, \pi_I(t,\, \bm{y})  \right] = i
 \delta^{(3)}({\bm x}- {\bm y})\,, \qquad \left[ \zeta_I(t,\, {\bm
 x}),\, \zeta_I(t,\, {\bm y}) 
 \right] =  \left[ \pi_I(t,\, {\bm x}),\, \pi_I(t,\, {\bm y}) \right] = 0\,. 
\end{align}

As we showed in Appendix \ref{Inin}, the $n$-point functions for
$\zeta$ given in Eq.~(\ref{Exp:zeta}) agree with those 
given by the solution of $\zeta$ written in terms of the retarded Green
function. We express the equation of motion for $\zeta$ schematically as
\begin{align}
 & \cL \zeta = {\cal S}_{\rm NL}\,, \label{eomt}
\end{align} 
where ${\cal L}$ denotes the differential operator
\begin{align}
 & {\cal L} := \partial^2_\rho + \left( 3 - \varepsilon_1+ \varepsilon_2 \right) \partial_\rho -
  \frac{\partial^2}{e^{2\rho}\dot{\rho}^2}\,,  \label{Def:cL}
\end{align}
and the left-hand side of Eq.~(\ref{eomt}) is the 
same equation of motion as is derived
from the non-interacting part of the action (\ref{Exp:S0}).

By using the retarded Green function:
\begin{align}
 & G_R(x,\, x') = i \theta(t-t') \left[ \zeta_I(x),\, \zeta_I(x') \right]\,,  \label{Def:GR}
\end{align}
which satisfies
\begin{align}
 & {\cal L} G_R(x, x') = - \frac{1}{2 \Mp^2} \frac{1}{\varepsilon_1
 e^{3\rho} \dot{\rho}^2} \delta^{(4)}(x-x') \,, \label{Eq:GR}
\end{align}
with
$
\delta^{(4)}(x-x'):= \delta(t - t')\delta^{(3)}(\bm{x}-\bm{x}')$,
the solution of the equation of motion with the initial condition
(\ref{IC}) is given by
\begin{align}
 & \zeta (x) = \zeta_I(x)  +  {\cal L}_R^{-1} {\cal S}_{\rm NL}(x) \label{SolR}
\end{align}
where the non-linear term is given by
\begin{align}
 & {\cal L}_R^{-1} {\cal S}_{\rm NL}(t,\, \bm{x})  := - 2 \Mp^2 \int^t_{t_i} \dd t' \!\int\! \dd^3
 \bm{x}'\varepsilon_1(t') e^{3\rho(t')} \dot{\rho}(t')^2 G_R(x, x') {\cal S}_{\rm NL}(x')\,. 
 \label{Exp:SolR}
\end{align}
Using Eq.~(\ref{SolR}), we can solve
the equation of motion iteratively.

We expand the interaction picture field $\zeta_I$, which satisfies
\begin{align}
 & {\cal L} \zeta_I(x)=0 \,, \label{Eq:GR}
\end{align}
as follows
\begin{equation}
  \zeta_I(x)= \int \frac{\dd^3 \bm{k}}{(2\pi)^{3/2}} \left(a_{\sbm{k}} v_k
   e^{i \sbm{k}\cdot \sbm{x}} + \mbox{h.c.} \right) . \label{Exp:expsi}
\end{equation}
The mode function $v_k$ satisfies ${\cal L}_k v_k=0$ where 
\begin{align}
 & {\cal L}_k := \partial^2_\rho + (3 - \varepsilon_1 + \varepsilon_2)
 \partial_\rho + \frac{e^{-2\rho}}{\dot{\rho}^2} k^2 \,.
\end{align}
The mode function is normalized as  
\begin{align}
 & \left( v_k e^{i \sbm{k} \cdot \sbm{x}},\, v_p e^{i \sbm{p} \cdot
 \sbm{x}} \right) = (2\pi)^3 \delta^{(3)} (\bm{k} - \bm{p})\,, \label{Cond:N}
\end{align}
where the Klein-Gordon inner product is defined by 
\begin{align}
  (f_1, f_2) := 
 - 2 i \Mp^2 \int \dd^3 \bm{x}\,  e^{3\rho} \varepsilon_1 
  \{ f_1 \partial_t f_2^* - (\partial_t f_1) f_2^*\}\,.
\end{align}
With this normalization, we obtain the commutation relations for the creation and
annihilation operators as
\begin{align}
 & \left[ a_{\sbm{k}},\, a_{\sbm{p}}^\dagger \right] =
 \delta^{(3)}(\bm{k}- \bm{p}), \qquad \left[ a_{\sbm{k}},\,
 a_{\sbm{p}} \right] = 0\,.
\end{align}
Inserting Eq.~(\ref{Exp:expsi}) into Eq.~(\ref{Def:GR}), we can rewrite
the retarded Green function as
\begin{align}
 & G_R(x,\, x') = i \theta(t-t') \!\int\! \frac{\dd^3 \bm{k}}{(2\pi)^3}
 e^{i \sbm{k} \cdot (\sbm{x} - \sbm{x}')} R_k(t,\, t')\,,
\end{align}
where $R_k(t, t')$ is given by
\begin{align}
 & R_k(t, t'):= v_k(t) v_k^*(t') - v^*_k(t) v_k(t')\,. \label{Exp:Rk}
\end{align}
We calculate the $n$-point function of $\gR$, setting 
the initial state to the vacuum defined by 
\begin{align}
 & a_{\sbm{k}} |0 \rangle =0 \,.  \label{Def:vacuum}
\end{align}

Next, we explicitly calculate non-linear corrections.
Since we are interested in the IR divergence, employing the
iteration scheme of ${\cal L}^{-1}_R$, we pick up only the terms
which can contribute to the IR divergences. For the convenience, we here introduce the symbol
``$\Approx$'' as in \cite{IRgauge} to denote an equality under the
neglect of the terms unrelated to the IR divergences. 
The IR divergences mean the appearance 
of the factor $\langle \zeta_I^2 \rangle$. For the scale invariant
spectrum, this variance diverges logarithmically. Once a temporal or
spatial differentiation acts on one of two $\zeta_I$s in $\langle \zeta_I^2 \rangle$, the variance no
longer diverges. In this sense we keep only terms which can yield 
$\langle \zeta_I^2 \rangle$. When we write down the 
Heisenberg operator $\zeta$ in terms of the interaction picture field
$\zeta_I$, at the one loop level,
this is equivalent to keep only the terms without 
temporal and/or spatial 
differentiations and the terms containing only one interaction picture 
field with differentiation.

In this paper, we refer to a term which
does (not) contribute to the IR divergences simply as an IR
(ir)relevant term. In the following, we will use
\begin{align}
  {\cal L}^{-1}_R \zeta_I {\cal R} \zeta_I
 \Approx \zeta_I {\cal L}^{-1}_R {\cal R} \zeta_I\,, \label{Prop1}
\end{align}
where ${\cal R}$ is a derivative operator which suppresses the IR modes
of $\zeta_I$ such as $\partial_\rho$ and
$\partial^i/\dot{\rho}e^{\rho}$.  Equation (\ref{Prop1}) can be
shown as follows. The Fourier transformation of 
${\cal L}^{-1}_R \zeta_I D \zeta_I$ is proportional to
\begin{align*}
 & \int \dd^3 \bm{p}\int^t_{t_i} \dd t' \varepsilon_1(t') e^{3\rho(t')} \dot{\rho}^2(t') R_{k}(t,\, t')
 \zeta_{I\,\sbm{p}}(t') \left( {\cal R} \zeta_I \right)_{\sbm{k}-\sbm{p}}(t')\,,
\end{align*} 
where $\zeta_{I\,\sbm{k}}$ and $({\cal R} \zeta_I)_{\sbm{k}}$ denote the Fourier
modes of $\zeta_I$ and ${\cal R} \zeta_I$. 
Since
$\left( {\cal R} \zeta_I \right)_{\sbm{k}-\sbm{p}}(t')-
\left( {\cal R} \zeta_I \right)_{\sbm{k}}(t')$ is suppressed and
$\zeta_{I, \sbm{p}}$ becomes time independent
in the limit $\bm{p} \to 0$, the IR relevant piece of the 
integrand of the momentum integral can be replaced by
\begin{align}
 & \zeta_{I\,\sbm{p}} \int^t_{t_i} \dd t' \varepsilon_1(t') e^{3\rho(t')}
 \dot{\rho}^2(t') R_{k}(t,\, t')
  \left({\cal R} \zeta_I \right)_{\sbm{k}}(t')\,.  \label{Prop3}
\end{align}
We will also use 
\begin{align}
 & {\cal L}^{-1}_R f(x) \Approx 0  \qquad \quad {\rm for}\,\, f(x)
 \Approx 0\,.  \label{Prop2}
\end{align}

Keeping only the IR relevant terms, the action for $\zeta$ is simply given by
\begin{align}
 &  S \Approx \Mp^2 \int  \dd t \dd^3 \bm{x} \,
e^{3(\rho+\zeta)}
 \dot{\rho}^2 \varepsilon_1
 \biggl[ (\partial_\rho \zeta)^2  -
 \frac{e^{-2(\rho+\zeta)}}{\dot{\rho}^2} \! ( \partial \zeta)^2  \biggr] \,,  \label{Exp:SIR_c}
\end{align} 
where we used the lapse function
$N$ and the shift vector $N_i$, given by solving the constraint
equations as follows,
\begin{align}
 & N \Approx 1 + \partial_\rho \zeta, \\
 & \frac{1}{\dot{\rho}} \partial_i N^i \Approx - \frac{e^{-2\rho}}{\dot{\rho}^2} \partial^2 \zeta + \varepsilon_1
 \partial_\rho \zeta\,.
\end{align} 
The action (\ref{Exp:SIR_c}) can be easily 
derived from the action for the non-interacting theory. 
Extending the action (\ref{Exp:S0}) to a nun-linear 
expression which preserves the dilatation symmetry, we obtain 
\begin{align}
 &  S = \Mp^2 \int  \dd t \dd^3 \bm{x} \,
e^{3(\rho+\zeta)}
 \dot{\rho}^2 \varepsilon_1
 \biggl[ (\partial_\rho \zeta)^2  -
 \frac{e^{-2(\rho+\zeta)}}{\dot{\rho}^2} \! ( \partial \zeta)^2 + \cdots  \biggr] \,,
\end{align} 
where the abbreviated terms do not 
appear in the non-interacting action. The abbreviated terms should have more
than two fields with differentiation. Such terms are 
IR irrelevant up to the one-loop order. Thus, we
obtain Eq.~(\ref{Exp:SIR_c}). 

Taking the variation of the action with respect to $\zeta$,
we obtain the evolution equation as
\begin{align}
 & \frac{1}{e^{3\rho}\dot{\rho} \varepsilon_1}
 \partial_\rho \left( e^{3\rho}
  \dot{\rho}\varepsilon_1 \partial_\rho \zeta \right) 
  - \frac{ e^{-2(\rho+\zeta)}}{\dot{\rho}^2} \partial^2 \zeta  \Approx  0\,.
\nonumber
\end{align}
By expanding $\zeta$ as 
$\zeta = \zeta_I + \zeta_2 + \zeta_3 \cdots$, the equation of motion is
recast into 
\begin{align}
 & {\cal L} \zeta_I = 0\,, \label{Eq:zeta1} \\
 & {\cal L} \zeta_2 \Approx -  2\zeta_I \Delta  \zeta_I\,, \label{Eq:zeta2}\\
 & {\cal L} \zeta_3 \Approx  - 2  
 \left( \zeta_2 \Delta \zeta_I + \zeta_I \Delta \zeta_2 - \zeta_I^2 \Delta \zeta_I
 \right)  \,, \label{Eq:zeta3}
\end{align}
where we defined
\begin{align}
 & \nabla_i:= \frac{e^{-\rho}}{\dot{\rho}} 
\partial_i\,, \quad  
 \Delta := \delta^{ij} \nabla_i \nabla_j\,.
\end{align}
Solving Eqs.~(\ref{Eq:zeta2}) and (\ref{Eq:zeta3}), we obtain 
\begin{align}
  &\zeta_2 \Approx  -  2 \zeta_I {\cal L}^{-1}_R \Delta \zeta_I\,,
 \label{Sol:zeta2} \\
 & \zeta_3 \Approx \zeta_I^2 {\cal L}_R^{-1} \left( 4 \Delta {\cal L}_R^{-1}
 \Delta + 2 \Delta \right) \zeta_I 
 \,, \label{Sol:zeta3}
\end{align}
where we noted the properties of ${\cal L}^{-1}_R$ given in
Eqs.~(\ref{Prop1}) and (\ref{Prop2}).

\subsection{Calculating the \gauge invariant operator}
Solving the three dimensional geodesic equations, we obtain the relation
between the global coordinates $x_{gl}^i$ and the geodesic normal coordinates
$x^i$ as
\begin{align}
 & x_{gl}^i \Approx e^{-\zeta(x)}  x^i\,.
\end{align}
Using the geodesic normal coordinates, the gauge-invariant curvature
$\gR$ is expressed as
\begin{align}
  &  \gR (t, \bm{x}) \Approx \sR(t,\, e^{- \zeta(t, e^{-\zeta}\sbm{x})} \bm{x})  \,.
\end{align}
 Here, we introduce the spatial average of $\zeta$ at the local
 observable
 region whose spatial scale is $L_t$ in the geodesic normal coordinates
 as
 \begin{align}
   & {^g\!\bar{\zeta}}(t) := \frac{\int \dd^3 \bm{x}\, W_{L_t}(\bm{x})
  \zeta(t, e^{-\zeta}\bm{x})
   }{\int \dd^3 \bm{x}\,W_{L_t}(\bm{x})}\,, 
\end{align} 
where $W_{L_t}(\bm{x})$ is the window function which specifies 
our observable region.
We approximate this averaging scale at each time $t$ by 
the horizon scale, {\it i.e.,} 
$L_t  \simeq \{e^{\rho(t)}\dot{\rho}(t)\}^{-1}$. 
Using ${^g\!\bar{\zeta}}(t)$, we decompose the spatial coordinates as   
\begin{align}
   \gR (t,\, \bm{x}) &\Approx \sR(t,\, e^{- \{ \zeta(t, e^{-\zeta}\sbm{x})
 - {^g\!\bar{\zeta}}(t) \}}  
e^{-   {^g\!\bar{\zeta}}(t)} \bm{x})   \,.  
\end{align}
Using the window function $W_{L_t}(\bm{x})$, we define a local average of an operator
$O(t,\, \bm{x})$ as 
\begin{align}
 & \bar{O}(t) :=  \frac{\int \dd^3 \bm{x} W_{L_t}(\bm{x}) O(t, \bm{x})
 }{\int \dd^3 \bm{x} W_{L_t}(\bm{x})}\,.
\end{align}
We note that the factor 
$e^{- \{ \zeta(t,e^{-\zeta}\sbm{x})- {^g\!\bar{\zeta}}(t)\}}$, which can be
expanded in perturbation as 
\begin{align}
 & e^{- \{ \zeta(t, e^{-\zeta}\sbm{x})-{^g\!\bar{\zeta}}(t)\}}  \cr 
 &\, = 1 - \{ (\zeta_I - \bar{\zeta}_I) + (\zeta_2 - \bar{\zeta}_2) - (\zeta_I \bm{x}
 \cdot \partial_{\sbm{x}} \zeta_I -\overline{\zeta_I \bm{x}
 \cdot \partial_{\sbm{x}} \zeta_I} \} + \frac{1}{2} (\zeta_I - \bar{\zeta}_I)^2 +
 \cdots\,,  
\end{align}
yields only IR regular contributions up to the one-loop order such as 
$\langle (\zeta_I - \bar{\zeta}_I) \zeta_I \rangle$ and 
$\langle \zeta_I \bm{x} \cdot \partial_{\sbm{x}} \zeta_I \rangle$. The
latter one also becomes finite because of the derivative suppression and the
finiteness of $|\bm{x}|$~\footnote{One may
think that interaction vertexes located outside the past lightcone can appear
from the inverse Laplacian $\partial^{-2}$ which are included in the
lapse function and shift vectors. In that case, $|\bm{x}|$ becomes unbounded.
However, tuning the boundary condition of the constraint
equations, which are the elliptic type equations, we can manifestly shut
off the influences from outside the observable region ${\cal O}$. Since
this change can be though of as the change of the residual \gauge degrees
of freedom, it does not influence on the \gauge invariant operator
$\gR$. The detailed explanation is given in Appendix of Ref.~\cite{SRV2}. }.
For the reason mentioned above, neglecting the factor 
$e^{- \{ \zeta(t,e^{-\zeta}\sbm{x})- {^g\!\bar{\zeta}}(t)\}}$, 
we rewrite the gauge invariant operator $\gR$ as
\begin{align}
   \gR (x) &\Approx \sR(t,\, e^{- {^g\!\bar{\zeta}}(t)} \bm{x})   \,.  
\end{align}

Since the spatial curvature $\sR$ is given by 
$$\sR(x) \Approx - 4 e^{-2\rho} e^{-2\zeta} \partial^2 \zeta\,, $$ using the
curvature perturbation in the geodesic normal coordinates:  
\begin{align}
 & \gz (t,\, \bm{x}) := \zeta (t,\,   e^{-{^g\!\bar{\zeta}}(t)} \bm{x})\,, \label{Exp:gz}
\end{align}
we can describe the gauge-invariant spatial curvature $\gR$ as
\begin{align}
 \gR(t,\, \bm{x}) &\Approx - 4 e^{-2\rho} e^{-2 (\zeta -
 {^g\!\bar{\zeta}})} \partial_{\sbm{x}}^2
 \zeta(t, e^{-{^g\!\bar{\zeta}}(t)}\bm{x}) \cr
 &\Approx - 4 e^{-2\rho} \partial_{\sbm{x}}^2 \gz(t,\, \bm{x})  \label{Rel:gRgz}
\end{align}
at least up to the third order in perturbation. At the second equality,
we again used the fact that the exponential factor
$e^{- (\zeta -{^g\bar{\zeta}})}$ do not
give IR relevant contributions. As is pointed out by Tsamis and Woodard in
Ref.~\cite{Tsamis:1989yu} and also in Ref.~\cite{Miao:2012xc}, using the
geodesic normal coordinates can introduce an additional origin of UV
divergence. We suppose that replacing $\zeta$ in the geodesic normal
coordinates with ${^g\bar{\zeta}}$, which is smoothed by the window
function, can moderate the singular behaviour.

Inserting the solution of $\zeta$
given in Eqs.~(\ref{Sol:zeta2}) and (\ref{Sol:zeta3}) into
Eq.~(\ref{Exp:gz}), we obtain
\begin{align}
 \gz_2  &\Approx - \bar{\zeta}_I \left( 2 {\cal L}_R^{-1} \Delta  +  \bm{x} \!\!\cdot\! \partial_{\sbm{x}} 
 \right) \zeta_I\,,  \label{Exp:gz2} \\
 \gz_3 
 & \Approx \frac{1}{2} \bar{\zeta}_I^2  \left( 2 {\cal L}_R^{-1} \Delta +  \bm{x}\!\!\cdot\! \partial_{\sbm{x}}
 \right)^2\zeta_I
+ \bar{\zeta}_I^2 (1- {\cal L}_R^{-1} {\cal L}) \bm{x}\!\cdot\!
 \partial_{\sbm{x}} {\cal L}_R^{-1} \Delta \zeta_I\,. \label{Exp:zeta3}
\end{align}
In deriving the expression of $\gz_3$, we used
\begin{align}
 & \left[ {\cal L},\, \bm{x} \cdot \partial_{\sbm{x}}  \right]  = - 2
 \Delta\,. \label{Exp:CommLX}
\end{align}
We can verify that the term with $(1- {\cal L}_R^{-1} {\cal L})$ in Eq.~(\ref{Exp:zeta3})
vanishes. Since this term is eliminated by operating $\cL$, it always vanishes 
if its value and its first time derivative are both zero  
at the initial time, which is automatically satisfied 
by the definition of the retarded integral. 

\subsection{Conditions for the absence of IR divergence}
We next discuss the condition that the IR divergence does not arise 
in the expectation values of the gauge invariant variable $\gR$. 
Then, using the above expression, two point function of $\gR$ up to the
one loop order is obtained as  
\begin{align}
  \langle \gR(x_1) \gR(x_2) \rangle
  &\Approx 8 e^{-4\rho}
  \langle \bar\zeta_I^2 \rangle 
 \partial^2_{(1)}  \partial^2_{(2)}  \Big\langle 2 
  \left( 2 {\cal L}_R^{-1} \Delta_{(1)} +  \bm{x}_1\!\!\cdot\! \partial_{\sbm{x}_1}
 \right)\! \zeta_I(x_1) 
 \left( 2 {\cal L}_R^{-1} \Delta_{(2)} +  \bm{x}_2\!\!\cdot\! \partial_{\sbm{x}_2} \right)\! \zeta_I(x_2)  
\cr &\qquad \qquad
 + \left( 2 {\cal L}_R^{-1} \Delta_{(1)} +  \bm{x}_1\!\!\cdot\! \partial_{\sbm{x}_1}
 \right)^2\!\! \zeta_I(x_1) \zeta_I(x_2) 
 + \zeta_I(x_1) \left( 2 {\cal L}_R^{-1} \Delta_{(2)} +  \bm{x}_2\!\!\cdot\! \partial_{\sbm{x}_2}
 \right)^2\!\! \zeta_I(x_2)  
 \Big\rangle, 
\label{DifR}
\end{align}
where we used an abbreviated notation 
$\partial^2_{(a)}:= \partial_{\sbm{x}_a}^2$ and 
$\Delta_{(a)} := (e^{\rho} \dot{\rho})^{-2} \partial^2_{\sbm{x}_a}$ for
$a=1,2$. As we mentioned at the end of the preceding section, the
correlation function of $\gR$ can 
contain IR divergences. One may think that the IR regularity is guaranteed if 
$( 2 {\cal L}_R^{-1} \Delta +  \bm{x}\!\!\cdot\! \partial_{\sbm{x}})\zeta_I=0$  
is imposed. However, this condition is in conflict with 
the use of the ordinary retarded integral in the iteration process.  
In fact, one can calculate 
$( 2 {\cal L}_R^{-1} \Delta +  \bm{x}\!\!\cdot\! \partial_{\sbm{x}}
)\zeta_I$ as 
\begin{align}
 &  \left( 2 {\cal L}_R^{-1} \Delta +  \bm{x}\!\!\cdot\! \partial_{\sbm{x}}
  \right)\zeta_I
  = \int{\dd^3 \bm{k} \over (2\pi)^{3/2}} \left\{
  e^{i\sbm{k}\cdot\sbm{x}} a_{\sbm{k}} 
 \left(-2
  {\cal L}_{R,k}^{-1} {k^2\over e^{2\rho}\dot\rho^2} +i\bm{x}\cdot\bm{k} \right) v_k + \mbox{h.c.}  \right\}\,,
\end{align}
where ${\cal L}^{-1}_{R,k}$ is the Fourier mode of ${\cal L}^{-1}_R$.
The requirement that this expression should identically vanish
leads to 
\begin{equation}
\left(-2  {\cal L}_{R,k}^{-1} {k^2\over e^{2\rho}\dot\rho^2} +i\bm{x}\cdot\bm{k} \right) v_k=0\,. 
\label{naivecondition}
\end{equation}
The first term is independent of $\bm{x}$, while the second one
manifestly depends on $\bm{x}$. This shows that the condition
$( 2 {\cal L}_R^{-1} \Delta +  \bm{x}\!\!\cdot\! \partial_{\sbm{x}})\zeta_I=0$
is incompatible with the use of the retarded integral.

In place of this naive condition, an alternative possibility one can think of is to impose 
\begin{equation}
 \left( 2 {\cal L}_R^{-1} \Delta +  \bm{x}\!\!\cdot\! \partial_{\sbm{x}}
  \right)\zeta_I =
\int{\dd^3 \bm{k} \over (2\pi)^{3/2}} \left(
  a_{\sbm{k}} D e^{i\sbm{k}\cdot\sbm{x}} v_k+\mbox{h.c.} \right)\, 
\label{bettercondition}
\end{equation}
with 
\begin{align}
 & D := k^{-3/2} e^{-i\phi(k)}
   \bm{k} \cdot \partial_{\sbm{k}} k^{3/2}e^{i\phi(k)}\, , 
\end{align}
where $\phi(k)$ is an arbitrary real function. 
This condition continues to hold once it and its time derivative 
are satisfied on a certain time slice because 
both sides of this equation 
vanish under an operation of the second 
order differential operator ${\cal L}$. 
The condition can be rewritten as a
condition on mode functions 
\begin{equation}
- 2  {\cal L}_{R,k}^{-1} { k^2\over e^{2\rho}\dot\rho^2} v_k=D v_k\,,
\label{Cond:GI}
\end{equation}
which is not contradictory as in the case of 
Eq.~(\ref{naivecondition}). With these conditions, 
the expectation value in Eq.~(\ref{DifR}) 
can be summarized in total derivative form as
\begin{align*}
 \langle \gR(x_1) \gR(x_2) \rangle
  \Approx 8 e^{-4\rho}
  \big\langle \bar\zeta_I^2 \rangle  \int \frac{\dd (\ln k) 
\dd \Omega_{\sbm{k}}
}{(2\pi)^3} \partial^2_{\ln k}
 \left\{ k^{7} |v_k|^2 e^{i \sbm{k}\cdot (\sbm{x}_1-\sbm{x}_2)}\right\}\,,
\end{align*} 
where we used 
$k^{3/2} D = e^{-i\phi(k)}
   \bm{k} \cdot \partial_{\sbm{k}} k^{3/2}e^{i\phi(k)}$ 
and $\int \dd \Omega_{\sbm{k}}$
denotes the
integration over the angular directions of $\bm{k}$. 
Since this integral of total derivative vanishes, the IR divergence disappears.

Although requesting the condition (\ref{Cond:GI}) can make the IR divergence
vanish, a bit more careful thought rules out this possibility. Since the
left hand side of  (\ref{Cond:GI}) and its time derivative vanish at the
initial time, the condition (\ref{Cond:GI}) requests
$D v_k(t_i)=D \dot{v}_k(t_i)=0$. Operating the differentiation
$\bm{k} \cdot \partial_{\sbm{k}}$ on the normalization condition of the mode functions,  
\begin{equation}
-2i e^{3\rho} \varepsilon_1 \left(
  v_k  \dot v^*_k
  -\dot v_k v_k^*\right)=1\, 
\end{equation}
leads to a contradiction. The right hand side
trivially vanishes after the operation of $\bm{k} \cdot \partial_{\sbm{k}}$, while the left
hand side gives 3.

Even though the IR regularity condition (\ref{Cond:GI}) cannot be
compatible with the initial conditions (\ref{IC}), it is still instructive to give
an alternative interpretation of the condition (\ref{Cond:GI}). 
We adopted the initial condition (\ref{IC}) for $\cv$, 
which identifies the Heisenberg fields with the corresponding 
interaction picture fields at the initial time and 
selected the vacuum state for the free field at the 
initial time. We denote a set of operations which specify the 
interacting quantum state by an
iteration scheme. In the canonical system of
$\cv$, we fixed the iteration scheme by Eq.~(\ref{IC}) and
Eq.~(\ref{Def:vacuum}).  
Then, when we take the same iteration scheme 
in the canonical system $\cvt$, 
it is not obvious whether the same vacuum state as the one in
the canonical system $\cv$ is picked up or not. 
Before closing this section, we show that the condition
(\ref{Cond:GI}) is identical to the condition that 
these two vacua are equivalent.

Since the transformation from $\cv$ to $\cvt$
is a canonical transformation, the correlation functions
for the same initial state calculated in these two
canonical systems should agree with each other, {\it i.e.},
\begin{align}
 & \langle \zeta(x_1) \zeta(x_2) \rangle 
 = \langle \tilde\zeta(t,\, e^s x^i_1) \tilde{\zeta}(t,\, e^s x^i_2) \rangle \,. \label{Exp:CT}
\end{align}
To employ the same iteration scheme, 
we request that both of $\zeta$ and
$\tilde{\zeta}$ are solved by using
${\cal L}^{-1}_R$ by identifying the Heisenberg fields 
with the interaction picture fields at the initial time. 
We expand the interaction picture
fields for $\zeta$ and $\tilde{\zeta}$, which we denote as $\zeta_I$ and
$\tilde{\zeta}_I$, respectively, in terms of the same mode function $v_k$ as
\begin{align}
 & \zeta_I(x) = \int \frac{\dd^3 \bm{k}}{(2\pi)^{3/2}} a_{\sbm{k}} v_k
 e^{i \sbm{k}\cdot \sbm{x}} + {\rm h.c.}\,,  \\
&  \tilde\zeta_I(x) = \int \frac{\dd^3 \bm{k}}{(2\pi)^{3/2}} \tilde{a}_{\sbm{k}} v_k
 e^{i \sbm{k}\cdot \sbm{x}} + {\rm h.c.}\,.   \label{Exp:tzI}
\end{align}
Further, we select the vacuum states that are erased under the 
action of $a_{\sbm{k}}$ and $\tilde{a}_{\sbm{k}}$ in the respective systems. 
We denote these vacuum states specified by $a_{\sbm{k}}$ and
 $\tilde{a}_{\sbm{k}}$ as $|\,0\, \rangle$ and 
$|\,\tilde{0}\, \rangle$, respectively. Now we show that
the equivalence between the two point functions, {\it i.e.},
\begin{align}
 & \langle\,0\,| \zeta(x_1) \zeta(x_2) |\,0\,\rangle 
 = \langle\,\tilde{0}\,| \tilde\zeta(t,\, e^s \bm{x}_1) \tilde{\zeta}(t,\, e^s \bm{x}_2)
 |\,\tilde{0}\,\rangle \,, \label{Exp:CT2}
\end{align}
yields the condition (\ref{Cond:GI}). We expand $\tilde\zeta(t,\, e^s x^i)$ as
\begin{align}
 \tilde{\zeta}(t,\, e^s \bm{x}) &=
 \tilde{\zeta}(x) + s \bm{x} \cdot \partial_{\sbm{x}} \tilde{\zeta}(x) +
 \frac{1}{2} s^2 (\bm{x} \cdot \partial_{\sbm{x}})^2 \tilde{\zeta} (x) +
 {\cal O}(s^3)  \cr
 & \Approx \tilde{\zeta}_I - \tilde\zeta_I 2 {\cal L}_R^{-1} \Delta \tilde\zeta_I 
+ s(2 {\cal L}_R^{-1} \Delta + \bm{x}\cdot \partial_{\sbm{x}}) \tilde\zeta_I +
 \cdots \,,
\end{align}
taking into account that the interaction
Hamiltonian is shifted by $-s$. Then, the right-hand side of Eq.~(\ref{Exp:CT2}) gives
\begin{align}
 &  \langle\,\tilde{0}\,| \tilde\zeta(t,\, e^s \bm{x}_1) \tilde{\zeta}(t,\, e^s \bm{x}_2)
 |\,\tilde{0}\,\rangle  \cr
 &\quad  \Approx  \langle\,\tilde{0}\,| \{ \tilde{\zeta}_I(x_1) - \tilde\zeta_I(x_1)
 2 {\cal L}_R^{-1}  \Delta_{(1)} \tilde\zeta_I(x_1) + \cdots  \} \{  \tilde{\zeta}_I(x_2) - \tilde\zeta_I(x_2)
 2 {\cal L}_R^{-1}  \Delta_{(2)} \tilde\zeta_I(x_2) + \cdots \}
 |\,\tilde{0}\,\rangle \cr
 & \qquad \quad  + s  \langle\,\tilde{0}\,| (2{\cal L}_R^{-1} \Delta_{(1)} + \bm{x}_1\cdot
 \partial_{\sbm{x}_1}) \tilde\zeta_I (x_1) \tilde{\zeta}_I(y) + \cdots  |\,\tilde{0}\,\rangle
 +  s  \langle\,\tilde{0}\,| \tilde\zeta_I (x)  (2 {\cal L}_R^{-1} \Delta + \bm{y}\cdot
 \partial_{\sbm{y}}) \tilde{\zeta}_I(y) + \cdots  |\,\tilde{0}\,\rangle \cr
 & \qquad \quad + {\cal
 O}(s^2)\,, \label{Exp:CT3}
\end{align}
where ``$\cdots$'' denotes higher order terms in perturbation.
Since the terms in the second line of Eq.~(\ref{Exp:CT3}) agree with
the left-hand side of Eq.~(\ref{Exp:CT2}), the other terms on the
right-hand side of Eq.~(\ref{Exp:CT3}) should vanish to satisfy
Eq.~(\ref{Exp:CT2}). The remaining terms of order $s$ at the leading
order in $\tilde{\zeta}_I$ in Eq.~(\ref{Exp:CT3}) is
given by 
\begin{align}
 & s \int \frac{\dd^3 \bm{k}}{(2\pi)^3} \left[ v_k^* e^{- i
 \sbm{k} \cdot \sbm{x}_2}  \{ 2 {\cal L}_R^{-1}\Delta_{(1)} + \bm{x}_1
 \cdot \partial_{\sbm{x}_1}\} v_k e^{i \sbm{k} \cdot \sbm{x}_1}
 +   v_k e^{i \sbm{k} \cdot \sbm{x}_1} \{ 2 {\cal L}_R^{-1}\Delta_{(2)} + \bm{x}_2
 \cdot \partial_{\sbm{x}_2}\} v_k^* e^{- i\sbm{k} \cdot \sbm{x}_2} \right]\, \cr
& = - s \int \frac{\dd^3 \bm{k}}{(2\pi)^3} e^{i \sbm{k} \cdot (\sbm{x}_1-\sbm{x}_2)} 
\left[ v_k^* 
 \left\{ 2 {\cal L}_{R,k}^{-1}\frac{k^2}{e^{2\rho}\dot{\rho}^2} v_k +k^{-3/2} \bm{k}
  \cdot \partial_{\sbm{k}} (k^{3/2} v_k) \right\}
 +  {\rm c.c.} \right]\,,
\end{align}
where we performed integration by parts. Now it is clear that 
requesting Eq.~(\ref{Exp:CT}) gives the same condition as requesting
the IR regularity (\ref{Cond:GI}), except for the irrelevant
$\bm{k}$-dependence of the phase of the mode functions.

\section{Summary and discussions}  \label{Sec:Con}

We have focused on observable quantities which are invariant under the
residual \gauge degrees of freedom. These degrees of freedom are left unfixed in the  
ordinary discussion of cosmological perturbation. 
The key issue for the correlation functions of the \gauge invariant operator to be IR regular
is shutting off the 
long range correlation between observable quantities and 
the fluctuation outside the observable region. 
This is the matter of how we choose the initial quantum state. 
The leading effect of the long range correlation 
on local observables is the constant shift of the so-called curvature 
perturbation ($=$ the trace part of spatial metric perturbation). 
The constant shift of the curvature perturbation can be 
absorbed by the dilatation transformation of the spatial coordinates. Assuming that the
interactions are shut off before the initial time, we investigated the
conditions that guarantee the equivalence between two systems mutually
related by the dilatation, and found that these conditions also guarantee
the IR regularity of observable quantities. Therefore, we can think of
the IR regularity condition (\ref{Cond:GI}) as requesting the invariance
under the dilatation, which is one of the residual \gauge
transformation.

We also found that these conditions are not compatible with 
the prescription where the vacuum state is set to the one 
for the corresponding non-interacting theory at an initial time. In this setup, the
initial time is very particular time, at which the Heisenberg picture fields
agree with the interaction picture fields. Then, the curvature scale at
the initial time becomes distinguishable from other scales. It is, therefore,
natural that the presence of the particular initial time breaks the
invariance under the scale transformation. One possibility to avoid
breaking the dilatation symmetry is sending the initial time to the 
infinite past. When we send the initial time to the past infinity,
the IR regularity no longer requests the condition (\ref{Cond:GI}),
because Eq.~(\ref{Prop1}) does not hold in this limit. At
Eq.~(\ref{Prop3}), we took $\zeta_{\sbm{p}}$ out of the time
integration, because $\zeta_{\sbm{p}}$ becomes constant in time in the
limit $p/e^{\rho}\dot{\rho} \ll 1$. However, since all the modes become the
sub Hubble modes at the distant past, the same argument does not
follow when we send the initial time to the past infinity. Thus, when we
keep the
interaction turned on from the past infinity, our claim in this paper
does not prohibit the presence of initial states which guarantee the IR
regularity.

Then, the question is whether there is an initial state (or an iteration
scheme) which guarantees the IR regularity/the \gauge invariance in the
local observable universe. In our previous
papers~\cite{IRgauge_L, IRgauge}, we found that when we specify the relation between the Heisenberg picture
field $\zeta$ and the interaction picture field $\zeta_I$ in a
non-trivial way and choose the adiabatic vacuum as the vacuum state for
the non-interacting theory, the IR contributions in the two-point functions of $\gR$
are regularized. The result of the present paper shows that the relation
between $\zeta$ and $\zeta_I$ imposed in Ref.~\cite{IRgauge_L,
IRgauge} is different from the relation fixed by the
initial condition (\ref{IC}), because the mode equation for the
adiabatic vacuum does not satisfy Eq.~(\ref{Cond:GI}).  We naturally
expect that the IR regular vacuum we found in Ref.~
\cite{IRgauge_L,IRgauge} corresponds to the iteration scheme where the
interaction has been active from the past infinity.

In general, when we keep the iteration turned on from the past
infinity, the time integration at each interaction vertex does not
converge. The $i \epsilon$ prescription provides a noble way to make the
time integration converge. The adiabatic vacuum is the vacuum state
which is selected by the $i\epsilon$ prescription. There is another advantage to fix the
iteration scheme by the $i\epsilon$ prescription. The
correspondence between the IR regularity and the \gauge invariance
provides an important clue to prove the IR regularity. Our result in the simple iteration scheme suggests that the
IR regularity of the loop corrections may be ensured, if we employ the
iteration scheme which satisfies
\begin{align}
 &  \langle \Omega \,| \zeta(x_1) \zeta_(x_2) \cdots \zeta(x_n) |\, \Omega
 \rangle =  \langle \tilde\Omega \,| \tilde\zeta(t_1,\, e^{-s}\bm{x}_1)
 \tilde\zeta(t_1,\, e^{-s}\bm{x}_2)  \cdots
 \tilde\zeta(t_1,\, e^{-s}\bm{x}_n)  |\, \tilde\Omega
 \rangle \,,  \label{condtion}
\end{align}
where  $|\,\Omega \rangle$ and $|\,\tilde\Omega \rangle$ are the initial
states selected by the iteration scheme in the two canonical systems
$\cv$ and $\cvt$, respectively. Since the
$i\epsilon$ prescription can be shown to select the unique state, which
becomes the ground state when the Hamiltonian is conserved in time, we
expect that the condition (\ref{condtion}) can be satisfied if we fix
the integration scheme by the $i\epsilon$ prescription. In our succeeding
paper~\cite{SRV2}, we will verify this expectation and will show that
the IR regularity and the absence of the secular growth can be ensured
if we employ the iteration scheme with the $i\epsilon$ prescription.

\acknowledgments
This work is supported by the Grant-in-Aid for the Global COE Program
"The Next Generation of Physics, Spun from Universality and Emergence"
from the Ministry of Education, Culture, Sports, Science and Technology
(MEXT) of Japan. T.~T. is supported by Monbukagakusho Grant-in-Aid for
Scientific Research Nos.~24103006, 24103001, 24111709, 21244033, 21111006.
Y.~U. is supported by the JSPS under Contact No.\ 21244033, MEC FPA under Contact No.\ 2007-66665-C02, and MICINN
project FPA under Contact No.\ 2009-20807-C02-02. Y.~U. would like to
thank J.~Garriga, I.~Khavkine, S.~Miao, R.~Woodard for their helpful
comments. 

\appendix 
\section{Solutions with the retarded Green function and the in-in
 formalism}  \label{Inin}
In this appendix, we discuss about the relation between the $n$-point
functions obtained from the solution written in terms of the retarded Green function $G_R$
and those obtained in the in-in formalism. We formally show that these $n$-point functions agree with each other. 
In the in-in formalism, the Heisenberg field $\zeta(x)$ is
related to the interaction picture field $\zeta_I(x)$ by the unitary
operator given by
\begin{align}
 & U_I(t_1,\, t_2) = T \exp \left[- i\int^{t_1}_{t_2} \dd t\, H_I(t) \right]\,,  
\end{align}
as follows
\begin{align}
 & \zeta (t, \bm{x}) =  U_I^\dagger(t,\,t_i)  \zeta_I(t, \bm{x}) 
U_I(t,\, t_i)\,. \label{Azeta}
\end{align}
Here, we keep the initial time $t_i$ finite.
Because of the unitarity of $U(t_1, t_2)$, the commutation
relation for $\zeta$ and its conjugate momentum $\pi_I$ automatically ensures
the commutation relation for $\zeta_I$ and its conjugate momentum
$\pi_I$. Using Eq.~(\ref{Azeta}), the $n$-point function for the
curvature perturbation $\zeta$ is given as   
\begin{align}
 \left\langle  \zeta(t, \bm{x}_1) \cdots \zeta(t, \bm{x}_n)
 \right\rangle  = \left\langle
 U_I^\dagger(t,\, t_i)  \zeta_I(t, \bm{x}_1)  \cdots  \zeta_I(t, \bm{x}_n)  
 U_I(t,\, t_i) \right\rangle.  \label{Exp:np} 
\end{align}
Using the mathematical induction, we can rewrite the $n$-point functions
(\ref{Exp:np}) as 
\begin{align}
  \left\langle \zeta(t, \bm{x}_1) \cdots \zeta(t, \bm{x}_n)
 \right\rangle&= \sum_{i=0}^\infty i^N \int^t_{t_i} \dd t_N \cdots \int^{t_2}_{t_i}
 \dd t_1 \cr
 & \quad \qquad \times \left\langle\!
  \Bigl[ H_I(t_1), \cdots \bigl[ H_I(t_N), \zeta_I(t, \bm{x}_1)  \cdots  \zeta_I(t, \bm{x}_n) 
  \bigr] \cdots \Bigr]\! \right\rangle.   \label{Exp:np2}
\end{align}
As a simple example, let us consider the two-point function in the case
where the action is given by
\begin{align}
 S &= 2 \Mp^2 \int \dd t\, \dd^3 \bm{X}\, e^{3\rho} \dot{\rho}^2
 \varepsilon_1 \left[ \frac{1}{2}(\partial_\rho \zeta)^2 -
 \frac{e^{-2\rho}}{2\dot{\rho}^2} (\partial \zeta)^2 - \frac{c_p}{p}
 \zeta^p \right]\,,  \label{Exp}
\end{align}
where $p$ is a natural number which is larger than 2 and $c_p$ is a
time-dependent function. In this case, the equation
of motion for $\zeta$ is simply given by 
\begin{align}
 & \cL \zeta  = c_p \zeta^{p-1} \,,
\end{align}
and the interaction Hamiltonian is given by
\begin{align}
 & H_I[\zeta_I] (t) = 2 \Mp^2  e^{3\rho} \dot{\rho}^2 \varepsilon_1
 \int \dd^3 \bm{x} \frac{c_p}{p} \zeta_I^p(t,\, \bm{x})\,.
\end{align}
Noticing the fact that the retarded Green function is expressed as
\begin{align}
 & G_R(x,\, x') = i \theta(t-t') \left[ \zeta_I(x),\, \zeta_I(x') \right], 
\end{align}
we first calculate the most inner commutation relation in
Eq.~(\ref{Exp:np2}) as
\begin{align}
 &i \int^t_{t_i} \dd t_N \bigl[ H_I(t_N), \zeta_I(t, \bm{x}_1)  \cdots  \zeta_I(t, \bm{x}_n) 
  \bigr] \cr
 &\,\, = - 2\Mp^2 \int^t_{t_i} \dd t_N \!\int\! \dd^3 \bm{x} e^{3\rho} \dot{\rho}^2 \varepsilon_1
 \frac{c_p}{p} \Bigl\{ {\cal C} \zeta_I^{p-1}(x_N) + \zeta_I(x_N) {\cal C}
 \zeta_I^{p-1}(x_N) + \cdots + \zeta_I^{p-1} (x_N) {\cal C} 
 \Bigr\}\,, \label{Sol:inin}
\end{align}
where we used the abbreviated notation $x_N:=(t_N,\, \bm{x})$ and 
$x_\alpha:= (t,\, \bm{x}_\alpha)$ for $\alpha=1,\cdots n$. We defined the operator ${\cal C}$ as 
\begin{align}
 {\cal C} &:= - i \theta(t- t_N) \left[ \zeta_I(x_N),\, \zeta_I(x_1) \cdots \zeta_I(x_n) \right] \cr
 &= \sum_{\alpha=1}^n G_R(x_\alpha, x_N) \prod_{m=1, m\neq \alpha}^n \zeta_I(x_m) \,.
\end{align}
Comparing this expression to Eqs.~(\ref{SolR}) and (\ref{Exp:SolR}), for
instance, the first term in the curly bracket of Eq.~(\ref{Sol:inin}) is
recast into
\begin{align}
 & \sum_{\alpha=1}^n \prod_{m=1, m\neq \alpha}^n \zeta_I(x_m) {\cal L}^{-1}_R \frac{c_p}{p}
 \zeta_I^{p-1} (x_\alpha)\,.
\end{align}
Repeating this procedure, we can show that the $n$-point functions
(\ref{Exp:np2}) agree with the $n$-point functions for $\zeta(x)$ which
is iteratively solved by using the retarded Green function. Here, for
illustrative purpose, we considered a simple case, but this argument can be
generalized in a straightforward manner.

\end{document}